# Spin-current-mediated rapid magnon localisation and coalescence after ultrafast optical pumping of ferrimagnetic alloys

E. Iacocca[1,2,3], T-M. Liu[4], A. H. Reid[4], Z. Fu[5], S. Ruta[6], P. W. Granitzka[4], E. Jal[4], S. Bonetti[4], A. X. Gray[4], C. E. Graves[4], R. Kukreja[4], Z. Chen[4], D. J. Higley[4], T. Chase[4], L. Le Guyader[4,7], K. Hirsch[4], H. Ohldag[4], W. F. Schlotter[4], G. L. Dakovski[4], G. Coslovich[4], M. C. Hoffmann[4], S. Carron[4], A. Tsukamoto[8], M. Savoini[9], A. Kirilyuk[9], A. V. Kimel[9], Th. Rasing[9], J. Stöhr[4], R. F. L. Evans[6], T. Ostler[10,11], R. W. Chantrell[6], M. A. Hoefer[1], T. J. Silva[2], H. A. Dürr[4,12]

1. Department of Applied Mathematics, University of Colorado, Boulder, CO 80309, USA
2. National Institute of Standards and Technology, Boulder, CO 80305, USA
3. Department of Physics, Division for Theoretical Physics, Chalmers University of Technology, Gothenburg 412 96, Sweden
4. SLAC National Accelerator Laboratory, 2575 Sand Hill Road, Menlo Park, CA 94025, USA
5. School of Physics, Science, and Engineering, Tongji University, Shanghai 200092, China
6. Department of Physics, University of York, York YO10 5DD, UK.
7. Spectroscopy & Coherent Scattering, European X-Ray Free-Electron Laser Facility GmbH, Holzkoppel 4, 22869 Schenefeld, Germany
8. Department of Electronics and Computer Science, Nihon University, 7-24-1 Narashino-dai Funabashi, Chiba 274-8501, Japan
9. Radboud University, Institute for Molecules and Materials, Heyendaalseweg 135, 6525 AJ Nijmegen, The Netherlands
10. Université de Liège, Physique des Matériaux et Nanostructures, Liège B-4000 Sart Tilman, Belgium
11. Faculty of Arts, Computing, Engineering and Sciences, Sheffield Hallam University, Howard Street, Sheffield, S1 1WB, UK
12. Department of Physics and Astronomy, Uppsala University, Box 516, 751 20 Uppsala, Sweden

**Sub-picosecond magnetisation manipulation via femtosecond optical pumping has attracted wide attention ever since its original discovery in 1996. However, the spatial evolution of the magnetisation is not yet well understood, in part due to the difficulty in experimentally probing such rapid dynamics. Here, we find evidence of rapid magnetic order recovery in materials with perpendicular magnetic anisotropy via nonlinear magnon processes. We identify both localisation and coalescence regimes, whereby localised magnetic textures nucleate and subsequently evolve in accordance with a power law formalism. Coalescence is observed for optical excitations both above and below the switching threshold. Simulations indicate that the ultrafast generation of noncollinear magnetisation via optical pumping establishes exchange-mediated spin currents with an equivalent 100% spin polarised charge current density of $10^8$ A/cm$^2$. Such large spin currents precipitate rapid recovery of magnetic order after optical pumping. These processes suggest an ultrafast optical route for the stabilization of desired meta-stable states, e.g., isolated skyrmions.**

Spin dynamics upon femtosecond optical pumping [1-15] have been intensely studied during the last two decades both because of potential applications for information storage and because of the need to understand the fundamental physics involved [16]. A variant of these dynamics is all-optical switching (AOS). While originally demonstrated for ferrimagnetic alloys with perpendicular magnetic anisotropy (PMA) [2], AOS has now been reported to occur in ferromagnetic PMA materials either subject to optical pumping [9, 10, 11, 12] or by use of ultrafast hot electrons [14, 15]. After ultrafast demagnetisation, the material's degrees of freedom can be considered to be in thermal equilibrium from the perspective of spatially averaged quasi-equilibrium dynamics, e.g., the three temperature model [1, 17, 18]. Whereas the three temperature model has been applied successfully to simulate picosecond magnetisation dynamics [1, 3, 4, 17] even to some degree for non-uniform states [5, 6] there is a growing understanding of the important role of spatially-varying magnetisation. For example, the chemical inhomogeneity of amorphous ferrimagnetic GdFeCo alloys results in picosecond transfer of angular momentum that both drives magnetisation switching [8] and influences the equilibrium state after pumping with a single laser pulse [13]. More recently, the effective domain size during cooling has been identified as a criterion to predict whether macroscopic AOS can occur [12].

To further investigate the fundamental physics involved in the evolution of spatially varying magnetisation after ultrafast optical pumping, and to elucidate which physical mechanisms are most important for the recovery of local magnetic order at picosecond timescales, we study the space- and time-dependent magnetisation dynamics in ferrimagnetic $Gd_{0.24}Fe_{0.665}Co_{0.095}$ alloys with time-resolved resonant X-ray scattering. We then compare our data with a multiscale model that utilizes both atomistic and large-scale micromagnetic components to simulate the time evolution of the magnetisation. We identify two distinct dynamic processes: magnon localisation and a subsequent magnon coalescence. These processes describe the nucleation and dynamics of localised textures that arise from nonlinear magnon interactions in contrast to an average, long-range magnetic order recovery associated to the thermalised magnon occupation distribution cool-down, e.g., as predicted in Ni [19].

Magnon localisation is the process by which a paramagnetic state evolves into a collection of localised spin textures, also known as magnon drops [20]. Our use of the term "magnon drop" in this case is topologically generic insofar as the spin texture in a single magnon drop is of an indeterminate winding number. Magnon localisation is characterised by the appearance of a ring pattern in the two-dimensional spin-spin correlation function that nucleates at the same length-scale as the microstructure of the magnetic material [8]. The subsequent magnon coalescence is characterized by the emergence of a broad peak in the spin-spin correlation function that is centred at low wavenumbers. This peak indicates weak long-range correlations in the spatial spin distribution and is a result of the continual nonlinear interaction of magnons that develop into randomly located magnon drops. By analysing our numerical simulations, an exchange flow spin current (EFSC) [21, 22] that is equivalent to a 100% polarised charge current density on the order of $10^8$ A/cm$^2$ is found. We propose that magnon drop perimeter deformations and dynamics driven by such spin currents expedites magnon coalescence via their growth, break-up, and merger.

Our study suggests that the picosecond evolution of the spatial magnetisation can be understood from a phase kinetics approach [23, 24]. In the case of nearly full demagnetisation upon femtosecond optical pumping, the system consists of a non-equilibrium distribution of randomised spins that then undergo rapid quenching of the magnetic order parameter that is subject to a possible multiplicity of equilibrium (or quasi-equilibrium) states. In other words, the subsequent rapid passage from a paramagnetic to a magnetically ordered state will generally do so via pathways of unstable domain growth, i.e., phase-ordering kinetics. Such dynamics contrast the critical behaviour expected from an adiabatic evolution through a phase transition [25]. Because of the possible degeneracy of the equilibrium, unstable growth necessarily leads to pattern formation, examples of which include domains in magnetic materials and metallic alloys [23], phase separation in binary fluids and superfluids [26], and optical solitons [27]. In addition, rapid quenching of the randomised state can dynamically stabilise topological defects via the Kibble-Zurek mechanism [28, 29], as seen in superfluids [26, 30], ferroelectrics [31], magnetic vortices [32], and bubble domain lattices [33]. Therefore, the phase kinetics interpretation of the magnon processes identified here sheds light onto the microscopic processes that must be controlled for macroscopic AOS or to stabilise desired equilibrium states upon ultrafast optical pumping.

The evolution of the spin-spin correlation function, $\Delta S_q^2$, is experimentally measured by time-resolved, coherent, resonant magnetic soft X-ray scattering, a pump-probe technique schematically shown in Figure 1**a** (see details in Methods). A 0.5 T field is applied perpendicular to the film plane during the measurement, such that the magnetisation is reset into the saturated state prior to optical pumping. The element-specific spatially-averaged dynamics are simultaneously measured by X-ray magnetic circular dichroism (XMCD) of the un-scattered beam. The scattering pattern provides information on the magnetisation's spatial profile. Two schematic examples are shown in Figure 1**b**. A ring in reciprocal space forms when there is a labyrinthine domain pattern in real space with a characteristic domain width, or correlation length, as shown in the top row. A broad peak centred at $q = 0$ forms when there are randomly located magnon drops, as shown in the bottom row.

We measured the magnetisation dynamics for both cases where the pump pulse fluence is below or above the AOS threshold. Sub-threshold dynamics were obtained with a 30 nm thick sample and an absorbed 800 nm pump fluence of 3.91 mJ/cm$^2$. In Figure 1**c**, the corresponding XMCD response for both Gd and Fe is constant between ≈3 ps and the longest delay time of 20 ps. AOS is obtained with a 20 nm thick sample and an absorbed 800 nm pump fluence of 4.39 mJ/cm$^2$. After switching, the XMCD data is also constant between 3 ps and 20 ps, as presented in Figure 1**d**. The slow time dependence of the XMCD data for both cases indicates that the average magnetisation is essentially constant for 3 ps $< t <$ 20 ps. A critical implication is that the quasi-thermal redistribution of magnon occupation caused by either damping or inelastic scattering that eventually drives the magnetisation towards a saturated state is not important at these timescales.

The azimuthally averaged spin-spin correlation function for Gd in the case of sub-threshold dynamics is shown by contours in Figure 1**e**. Spin-spin correlation profiles at selected time instances are shown in Figure 1**f** by solid black curves that have been shifted vertically for clarity. These lineouts have two spectral features; one centred close to or below the smallest resolved wavenumbers and one centred in the range 0.4 nm$^{-1}$ $< q <$ 0.8 nm$^{-1}$. Fits to the data shown by the dashed red curves are obtained by using a Gaussian line shape for the high-$q$ feature (with a peak position indicated by black circles) and a Lorentzian line shape for the low $q$ peak. The fitted Gaussian line shape indicates the appearance of a ring and therefore suggests the formation of a spatially correlated magnetisation pattern at sub-picosecond timescales. After ≈ 5 ps, reliable fits were obtained by use of only a Lorentzian line shape. Because the XMCD data remains quenched for the measurement time, we conclude that the Lorentzian feature corresponds to randomly located magnon drops [20].

For the case where the pump was sufficient to induce AOS, the azimuthally averaged spin-spin correlation shown in Figure 1**g** exhibits a peak at low $q$ that appears in a fraction of a picosecond. In this measurement, the maximum measured wavenumber of $q \approx 0.46$ nm$^{-1}$ was insufficient to determine the appearance of a Gaussian peak at higher wavenumbers. Spin-spin correlation profiles at selected time instances are shown in Figure 1**h**. Again, the curves are shifted vertically for the sake of clarity. Reliable fits were obtained solely by use of a Lorentzian peak, as shown with the dashed red curves in Figure 1**h**. As in the sub-threshold case, this spectral feature is consistent with that expected for a randomly located collection of magnon drops and suggests that macroscopic AOS at equilibrium requires magnon drops to merge into a single domain.

We performed atomistic simulations [34, 35] to understand better the physical mechanisms that are most important in driving the evolution of the spin-spin correlation dynamics after pumping. The amorphous alloy is modelled as a polycrystalline Gd and Fe-Co thin film with elemental inhomogeneity with a characteristic length of 7 nm, guided by recent experimental results [8]. The spatially averaged magnetic moments for Gd and Fe obtained with atomistic simulations are shown in Figure 2**a** for the case of sub-threshold dynamics utilising an absorbed fluence of 10.7 mJ/cm$^2$ and Figure 2**b** for the case of AOS utilising a very similar absorbed fluence of 11 mJ/cm$^2$. The atomistic simulations assume uniform heating across the thickness, and the utilised fluences are tuned to qualitatively reproduce the experimental XMCD data, cf. to Figure 1**c** and **d**. Snapshots of the simulated spatial magnetisation evolution are shown in Figure 2**c** and **d** for sub-threshold dynamics and AOS. In both cases, the coarsening of the spatially varying perpendicular-to-plane magnetisation from a fine-grained randomised state into a collection of magnon drops is observed. Such coarsening in the magnetic texture at such short time-scales is necessarily the result of non-conservative nonlinear magnon interactions, whereby spatial localisation rapidly minimizes magnon energy [20, 36]. This is in contrast to a simple picture of the field-driven growth of domains in an applied field, as is expected to be operative on much longer timescales greater than hundreds of picoseconds [37].

To directly compare with the experimental results, the simulated spin-spin correlation function is calculated via Fourier analysis of the spatially-dependent perpendicular-to-plane magnetisation. Contours of the azimuthally averaged spin-spin correlation function are shown in Figure 3**a**. Lineouts at selected time instances are shown in Figure 3**b** in addition to fits by a linear combination of a Lorentzian and a Gaussian centred at $q > 0$ with peak positions indicated by black circles. While the appearance of the Gaussian peak is less apparent than in the case for the data in Figure 1**f**, the fitting was unambiguous, as we further demonstrate below. For the case of AOS, contours of the azimuthally averaged spin-spin correlation function are shown in Figure 3**c** while selected lineouts and Lorentzian fits are shown in Figure 3**d** by solid black and dashed red curves, respectively. Both cases qualitative agree with the experimental data.

To further identify the role of exchange coupling between the rare earth and transition metal lattices, we performed multiscale micromagnetic simulations based on the Landau-Lifshitz (LL) equation [38] that consider an effective, homogeneous exchange stiffness. The ferrimagnetic GdFeCo is modelled as a single-species ferromagnet, with an initial condition provided by the atomistic simulations at a specified time $t_c \geq 3$ ps after optical pumping. By use of this multiscale approach, we can isolate the role of the atomic-scale exchange interactions, which dominate at short times, from the longer-range exchange stiffness. The choice of $t_c$ has a negligible effect on the qualitative features of the simulation results (see SI). As such, we only show a representative example at $t_c = 3$ ps.

For the sub-threshold case, the azimuthally averaged spin-spin correlation function is shown in Figure 3**e**. The black area indicates the temporal range in which atomistic simulations are used to calculate the initial conditions for the micromagnetic simulations. Corresponding lineouts, along with fits by the previously described sum of Lorentzian and Gaussian functions, are shown in Figure 3**f** by, respectively, solid black and dashed red curves. A striking feature in the micromagnetic simulations is the appearance of an additional Gaussian peak with a centre position identified by black circles in Figure 3**f**. This peak suggests the emergence of a material-independent natural correlation length in the magnetisation distribution even in the absence of chemical inhomogeneity. The general mechanism for the nucleation of such a correlation length is modulational instability, whereby magnons are localised by the nonlinear attractive potential driven by uniaxial anisotropy [20, 36, 39]. After 10 ps, only the Lorentzian component can be reliably fitted to the micromagnetic results. For the case of AOS, the azimuthally averaged spin-spin correlation function is shown in Figure 3**g**. Lineouts and corresponding Lorentzian fits are shown in Figure 3**h**. The qualitative agreement to both experimental data and atomistic simulations indicates that atomic-scale exchange interactions have a limited influence on the dynamics when only a Lorentzian line shape can be fitted.

To conclusively elucidate the physical mechanisms that drive the magnetisation dynamics within 20 ps after the optical pulse, we analyse the fitted parameters obtained from experiments and numerical simulations. We first study the Gaussian line shape observed during sub-threshold dynamics. The fitted peak position, $q_{max}$, and peak width, $\sigma_q$, of the Gaussian feature as a function of time are shown in Figure 4**a** and **b**, respectively. The blue circles are obtained from fits to experiments. For the first $\approx 3$ ps, both the central position and the peak width are approximately constant at $q_{max} = 0.57 \pm 0.014$ nm$^{-1}$ and $\sigma_q = 0.24 \pm 0.002$ nm$^{-1}$, respectively, indicating that the perpendicular magnetisation component pattern has a characteristic correlation length of $2\pi / 0.57$ nm$^{-1} \approx 11$ nm. Because the peak width is relatively constant, we conclude that the pattern is seeded by a static structure in the system, i.e., chemical inhomogeneities. This conclusion is in agreement with the ultrafast angular momentum transfer between regions of $\approx 10$ nm average chemical correlation length in similar amorphous GdFeCo alloys [8]. The red circles in Figure 4**a** and **b** are obtained from the atomistic simulations. The small error bars indicate that the fit was unambiguous. The average centre position estimated from the first 2 ps is $q_{max} = 0.88 \pm 0.012$ nm$^{-1}$ and it is accompanied by an approximately constant peak width at $\sigma_q = 0.18 \pm 0.04$ nm$^{-1}$. The corresponding length scale of $\approx 7.1$ nm agrees with the modelled correlation length of the chemical inhomogeneity. These simulations demonstrate how the chemical nonuniformity of the alloy seeds the magnetisation pattern within less than 1 ps. Between $\approx 3$ ps and $\approx 4.5$ ps the centre position from fits to experiments shifts towards $q = 0$ while fits to atomistic simulations are unreliable due to the large error bars. These concurrent observations indicate that the magnetic system dissociates from the sample's chemical inhomogeneity.

The nucleation of a correlated state with a concomitant diffraction ring and its subsequent dissociation evidenced by a transition into a central peak, defines the magnon localisation process.

The subsequent evolution of a collection of uncorrelated magnon drops is quantified from the Lorentzian fits to the central peak. The linewidth of the central peak provides information on the magnon drops and consists of four contributions: mean diameter, mean perimeter profile, and the statistical distribution of both. The influence of both the perimeter profile and distribution can be neglected based on intrinsic scale separation, i.e., the perimeter length scale is inversely proportional to the magnon drop diameter [20]. However, it is difficult to disentangle the mean magnon drop diameter from its distribution, partly due to the limited statistics that can be accumulated at picosecond timescales. Here, we will consider that the central peak full-width at half-maximum, $\Delta q$, provides a metric for the temporal evolution of magnon drops, calculated as a characteristic length scale $L(t) = 2\pi / \Delta q(t)$. In other words, an increase in $L(t)$ indicates either magnon drop growth, an increased uniformity in the magnon drops' size distribution, or some combination of both.

The characteristic length scales obtained from experiments, atomistic simulations, and micromagnetic simulations in the case of sub-threshold dynamics are shown in Figure 4**c**, by blue, red, and black circles, respectively. The experimental data indicates that the characteristic length scale is approximately 10 nm at 4 ps, which then grows to 50 nm at 20 ps. The rate of the domain growth closely follows a power law from 8 to 20 ps. Atomistic simulations are in quantitative agreement with the experimental results. Micromagnetic simulations exhibit a delay in comparison to the data and the atomistic simulations, but the growth rate is proportionally in good qualitative agreement with both. We attribute this disagreement to the spatial smoothing introduced by the continuum approximation and the different mechanism that drives the correlation lengths before 10 ps, i.e., modulational instability as opposed to chemical inhomogeneities.

The regime in which the characteristic length scale grows according to a power law is referred to as magnon coalescence. Because micromagnetic simulations exhibit a similar power law growth, we conclude that the combination of uniaxial anisotropy and exchange drives the process of magnon coalescence. This process is necessarily nonlinear as the balance between uniaxial anisotropy and exchange favours magnon drops as dynamical magnon bound states.

For the case of AOS, power law growth of the characteristic length scale is also obtained, as shown in Figure 4**d**. Before 7 ps, the experimental results exhibit a plateau corresponding to correlation length scales of $\approx$ 80 nm. From the XMCD data in Figure 1**d**, the magnetic moments are dynamically quenched for the first $\approx$ 3 ps so that the nucleation of magnon drops in this temporal range is probably hampered by the lack of any net magnetization to break the symmetry. Therefore, the plateau originates from scattering whose physical origin is unclear. In contrast, the atomistic simulations exhibit rapid growth between 2 ps and $\approx$ 5 ps, as indicated in the gold-shaded area. However, after this rapid localisation process, there follows much slower growth, indicated in the blue-shaded area, that is in good qualitative agreement with the experimental results shown by blue circles. Power law growth is also obtained from micromagnetic simulations, exhibiting a similar delay as the sub-threshold dynamics.

Power law fits are shown in Figure 4**c** and **d** by colour coded dashed lines that utilise the fitting function $L(t) = bt^a$ with the resultant fitting parameters listed in Table 1. We find exponents in the range $0.57 < a < 1.08$ for all cases. Similar analysis from experimental data obtained at different fluences for both Gd and Fe return exponents in the same range of values (see SI). Taking into account exponents obtained from experiments and simulations, an average exponent of $a = 0.85 \pm 0.1$ is found. For comparison, the Lifshitz-Cahn-Allen theory for the power law growth of uniaxial domains predicts an exponent of 1/2 [23, 40, 41]. However, the Lifshitz-Cahn-Allen theory assumes locally equilibrated domains, i.e., at long times when the magnon drop's perimeter dynamics is neglected. We conjecture that the faster growth rate observed here is the result of the non-equilibrium magnetisation dynamics that are present at short times.

Dynamical magnon drop perimeter deformations can favour fast magnon drop growth. The torque exerted by non-collinear spins in the form of EFSCs [21, 22, 42] can drive perimeter deformations of magnon drops. The EFSC density expressed as an equivalent 100% polarised charge current density, $\mathbf{J}_s$, can be numerically calculated from the magnetisation vector via a hydrodynamic representation [21, 43]. The normalised probability distribution for the EFSC density magnitude, $P(\mathbf{J}_s)$, at selected time instances is shown in Figure 4**e** and **f** for sub-threshold dynamics and AOS, respectively. Current densities on the order of $10^8$ A/cm$^2$ persist well after ultrafast pumping. For comparison, current densities of $\approx 10^7$ A/cm$^2$ are typical of those used to drive magnetisation switching via spin transfer torque [44]. Such large spin currents spatially deform the magnon drops' perimeters, establishing a multitude of modes that may include breathing and rotation [45]. In the dynamics studied here, such modes increase interactions between magnon drops that result in both merging and break-up [46]. Examples of magnon drop merging and break-up from micromagnetic simulations are shown in Figure 4**g**. Snapshots spanning 2 ps are shown. The magnon drops' perimeters where the perpendicular-to-plane magnetization is zero are shown in solid black areas. The gray and white areas indicate that the perpendicular-to-plane magnetisation is parallel or anti-parallel to the applied field. The curves represent the flow of EFSCs that transfer perpendicular-to-plane angular momentum, colour coded by the current density magnitude. Merging between the leftmost and central magnon drops is caused by strong EFSC flows that transfer angular momentum between the magnon drops. Break-up is observed at the top of the central magnon drop, where the EFSC flow transfers angular momentum away from the magnon drop. We also note that the EFSC flow exhibits curved trajectories, which suggests the existence of local magnetic topological defects that may eventually annihilate or stabilise long-lived topological textures, e.g., skyrmions.

Our results suggest that desired magnetisation states may be stabilised by nanopatterning magnetic materials to take advantage of both sub-picosecond seeded magnetisation states and EFSCs. For example, a close-packed spatially periodic pattern is expected to favour magnon coalescence and lead to a fast, macroscopic AOS; while engineered defects may lead to the stabilisation of isolated magnetic skyrmions via the Kibble-Zurek mechanism at picosecond timescales in materials with Dzyaloshinskii-Moriya interaction.

**Methods**

*Experiments*

The GdFeCo samples were fabricated on 100 nm thick Si$_3$N$_4$ membranes by magnetron sputtering. A 5 nm seed layer of Si$_3$N$_4$ was first grown on the membrane followed by the Gd$_{0.24}$Fe$_{0.665}$Co$_{0.095}$ film, which was then capped with 20 nm of Si$_3$N$_4$. X-ray measurements were conducted at the SXR hutch of the Linac Coherent Light Source [47]. The X-ray energy was selected to be resonant with the Fe L$_3$ resonance edge at 707 eV or the Gd M$_5$ resonance edge at 1185 eV with a 0.5 eV bandwidth and a pulse duration of 80 fs. The X-ray pulses were circularly polarised at the Fe $L_3$ and Gd $M_5$ edges by using the XMCD in magnetized Fe and GdFe films respectively placed upstream of the experiment. A degree of polarization was 85% at the Fe $L_3$ edge and 79% at the Gd $M_5$ edge. Measurements were made in transmission geometry with X-rays incident along the sample normal. An in-vacuum electromagnet was used to apply a field of 0.5 T perpendicular to the GdFeCo film. The diffracted X-rays were collected with a p-n charge-coupled device (pnCCD) two-dimensional detector placed behind the sample. A hole in the centre of the detector allowed the transmitted beam to propagate to a second detector used to collect the transmitted X-ray beam. The experiment was conducted in an optical pump – X-ray probe geometry. Optical pulses of 1.55 eV and 50 fs duration were incident on the sample in a near collinear geometry. The delay between the optical and X-ray pulses was achieved using a mechanical delay line, where the delay was continuously varied. X-ray–optical jitter was monitored and removed from the experimental data using an upstream cross-correlation arrival monitor [48].

*Atomistic simulations*

A model system of a GdFe ferrimagnet was developed to perform numerical simulations of the atomistic spin dynamics after femtosecond laser excitation. The inhomogeneous microstructure is generated by specifying random seed points representing areas of segregation of the Gd from the alloy, leading to 15% to 30% higher local Gd concentration. These regions are interpolated using a Gaussian with a standard deviation of 5 nm, representing the scale of the segregation. Due to low packing of the seed points, the characteristic length of the spatial variations is approximately 7 nm. An atomistic level simulation model is used to properly describe the ferrimagnetic ordering of the atomic moments with Heisenberg exchange [34]. The energy of the system is described by the spin Hamiltonian

$$\mathcal{H} = -\sum_{i<j} J_{ij} \mathbf{S}_i \cdot \mathbf{S}_j - \sum_i k_u (S_i^z)^2, \qquad (1)$$

where the spin $\mathbf{S}_i$ is a unit vector describing the local spin direction. $J_{ij}$ is the exchange integral, which we limit to nearest neighbour interactions. $k_u$ is the anisotropy constant and $\mu_s$ is the local (atomic) spin magnetic moment. Time-dependent spin dynamics is governed by the Landau-Lifshitz-Gilbert (LLG) equation at atomistic level

$$\partial_t \mathbf{S}_i = -\frac{\gamma}{(1+\alpha^2)} [\mathbf{S}_i \times \mathbf{B}_{\text{eff}}^i + \alpha \mathbf{S}_i \times (\mathbf{S}_i \times \mathbf{B}_{\text{eff}}^i)], \qquad (2)$$

where $\gamma$ is the gyromagnetic ratio and $\alpha = 0.01$ is the Gilbert damping factor. The on-site effective induction can be derived from the spin Hamiltonian with the local field augmented by a random field to model the interactions between the spin and the heat bath

$$\mathbf{B}_{\text{eff}}^i = -\frac{\partial \mathcal{H}}{\partial \mathbf{S}_i} + \varsigma_i, \qquad (3)$$

where the second term $\varsigma_i$ is a stochastic thermal field due to the interaction of the conduction electrons with the local spins. The stochastic thermal field is assumed to have Gaussian statistics and satisfies

$$\langle \varsigma_{i,a}(t) \varsigma_{j,b}(t') \rangle = \delta_{ij} \delta_{ab} (t-t') 2\alpha_i k_B T_e \mu_i / \gamma_i, \qquad (4)$$

$$\langle \varsigma_{i,a}(t) \rangle = 0, \tag{5}$$

where $k_B$ is the Boltzmann constant and $T$ is the temperature. We incorporate the rapid change in thermal energy of a system under the influence of a femtosecond laser pulse. The spin system is coupled to the electron temperature, $T_e$, which is calculated using the two-temperature model [49] with the free electron approximation for the electrons

$$C_e \frac{dT_e}{dt} = -G_{el}(T_l - T_e) + P(t), \tag{6}$$

$$C_l \frac{dT_l}{dt} = -G_{el}(T_e - T_l), \tag{7}$$

where $C_e = 225$ J m$^{-3}$ K$^{-1}$, $C_l = 3.1 \times 10^6$ J m$^{-3}$ K$^{-1}$, $G_{el} = 2.5 \times 10^{17}$ W m$^{-3}$ K$^{-1}$, and $P(t)$ models the temperature from a single Gaussian pulse into the electronic system. The pulse has a width of 50 fs.

We use Heun numerical integration scheme to integrate the stochastic equation of motion with time-varying temperature [35]. We use $\mu_{Fe} = 1.92\mu_B$ as an effective magnetic moment containing the contribution of Fe and Co and we set $\mu_{Gd} = 7.63\mu_B$ for the Gd sites. The standard parameters of the exchange coupling constants are used: $J_{\text{Fe-Fe}} = 4.526 \times 10^{-21}$ J per link, $J_{\text{Gd-Gd}} = 1.26 \times 10^{-21}$ J per link, and $J_{\text{Fe-Gd}} = -1.09 \times 10^{-21}$ J per link. We assume a uniaxial anisotropy energy of $8.07246 \times 10^{-24}$ J per atom. The numerical simulations are conducted using the VAMPIRE software package [35].

*Multiscale micromagnetic simulations*

Micromagnetic simulations were performed with the graphic processing unit (GPU) package MuMax3 [50] that solves the Landau-Lifshitz equation for a ferromagnet

$$\partial_t \mathbf{m} = -\gamma \mu_0 [\mathbf{m} \times \mathbf{B}_{\text{eff}} + \alpha \mathbf{m} \times \mathbf{m} \times \mathbf{B}_{\text{eff}}], \tag{8}$$

where **m** is the magnetisation vector normalised to the saturation magnetisation and $\mathbf{B}_{\text{eff}}$ is an effective induction that includes the required physical terms to model a ferromagnetic material. Here, we included exchange, non-local dipole, uniaxial anisotropy, and external fields. The exchange interaction in the micromagnetic approximation takes the form of a Laplacian scaled by the exchange length, $\lambda_{\text{ex}}$. In MuMax3, the Laplacian is numerically resolved by a 4$^{\text{th}}$ order central finite difference scheme, i.e., each micromagnetic cell is subject to exchange interaction due to itself and two neighbouring cells in each dimension. This approach offers numerical stability but also results in the smoothing of the magnetisation in cubic volumes of side length of 5$\lambda_{\text{ex}}$ that quickly supresses the short-range spin randomisation that occurs in atomistic simulations. This spatial smoothing leads to a temporal delay in the evolution of the spatially varying magnetisation. We ran our simulations on NVIDIA GPU units K20M, K40, K80, and P100. Due to the coarse resolution of micromagnetic simulations, we utilise approximately cubic cells of size 2 nm x 2 nm x $\delta$, where $\delta = 2^N D$ and the factor $N$ is chosen to take advantage of the GPU spectral calculations such that $\delta < \lambda_{\text{ex}} \approx 5$ nm and D is the physical thicknesses equal to 30 nm or 20 nm for the sub-threshold or switching dynamical cases, respectively. Note that the size of the cells only impacts the stability and accuracy of the numerical algorithm while the physics can only be interpreted in the framework of the continuum Landau-Lifshitz equation, i.e., long-wavelength features relative to the exchange length. We set the software to solve equation (8) with an adaptive-step, 4th order Runge-Kutta time integration method. Periodic boundary conditions (PBCs) were imposed along the film's plane. For both dynamical behaviours we used the equilibrium magnetic parameters: $M_S = 47170.6$ A/m, anisotropy constant $k_u = 31127.228$ J/m$^2$, exchange constant $A = 1$ pJ/m, and $\alpha = 0.01$. The value for $A$ was numerically found to best match the atomistic, average perpendicular magnetisation (See SI).

*Estimation of the change in the spin-spin correlation function*

Experimentally, the change in the spin-spin correlation function, $\Delta S_q^2$, was obtained from the scattered intensities measured by circularly polarised light as

$$\Delta S_q^2 = \frac{I_+(q,t)+I_-(q,t)}{2} - \langle\frac{I_+(q,t<0)+I_-(q,t<0)}{2}\rangle, \tag{9}$$

where $I_+(q,t)$ and $I_-(q,t)$ are the time-dependent scattered intensities obtained with right-handed and left-handed circularly polarised light. The background was subtracted by averaging the data collected at times before the optical pulse irradiated the sample.

The spin-spin correlation function for both atomistic and micromagnetic simulations was estimated by computing a two-dimensional fast Fourier transform (FFT) on the perpendicular magnetisation for each layer as a function of time. To minimise error, the FFTs obtained for each layer at a given time were averaged. No window function was used due to the PBCs.

*Equilibrium spin currents established by noncollinear magnetisation*

In the dispersive hydrodynamic formulation of magnetisation dynamics [21], the normalised magnetisation vector $\mathbf{m} = (m_x, m_y, m_z)$ in equation (8) can be cast in hydrodynamic variables by the canonical transformation

$$n = m_z \quad , \quad \mathbf{u} = -\nabla \arctan[m_y/m_x], \tag{10}$$

where $n$ is the density and $\mathbf{u}$ is the fluid velocity. For the case of conservative dynamics, $\alpha = 0$ in equation (8), the dispersive hydrodynamic equations are

$$\partial_t n = \nabla \cdot [(1 - n^2)\mathbf{u}], \tag{11}$$

$$\partial_t u = -\nabla[(1 - |\mathbf{u}|^2)n] - \nabla\left[\frac{\Delta n}{1-n^2} + \frac{n|\nabla n|^2}{(1-n^2)^2}\right] - \nabla h_0, \tag{12}$$

expressed in dimensionless space, time, and field scaled by, respectively $\sqrt{|H_k/M_s - 1|}\lambda_{\text{ex}}^{-1}$, $\gamma\mu_0|H_k - M_s|$, and $M_s^{-1}$, where the anisotropy field is given by $H_k = 2\,k_u/(\mu_o M_s)$, and $h_0$ is a dimensionless field applied normal to the plane. The longitudinal spin density flux in equation (4) is identified as the EFSC in hydrodynamic variables. To establish a clear comparison to spin currents obtained by charge-to-spin transduction, the EFSC are expressed as a 100% spin polarised charge current density in units of A/m$^2$ by [22]

$$\mathbf{J}_s = -\frac{2e}{\hbar}\mu_0 M_s^2 \lambda_{\text{ex}}(1 - n^2)\mathbf{u}. \tag{13}$$

We note that the factor $(1 - n^2)$ leads to maximum EFSC for a given **u** when the magnetisation is in the plane. For this reason, the magnon drop perimeters are primarily subject to EFSCs.

**Acknowledgements**

This material is based upon work supported by the U.S. Department of Energy, Office of Science, Office of Basic Energy Sciences under Award Number 0000231415 and is partly supported by the European Research Council ERC Grant agreement No. 339813 (Exchange) and the Netherlands Organisation for Scientific Research (NWO). Operation of LCLS is supported by the U.S. Department of Energy, Office of Basic Energy Sciences under contract No. DE-AC02-76SF00515. This work used the ARCHER UK National Supercomputing Service (http://www.archer.ac.uk). This project has received funding from the European Union's Horizon 2020 research and innovation programme under grant agreement No. 737093 (FEMTOTERABYTE). This work was performed using resources provided by the Cambridge Service for Data Driven Discovery (CSD3) operated by the University of Cambridge Research Computing Service (http://www.csd3.cam.ac.uk/), provided by Dell EMC and Intel using Tier-2 funding from the Engineering and Physical Sciences Research Council (capital grant EP/P020259/1), and DiRAC funding from the Science and Technology Facilities Council (www.dirac.ac.uk). E.I. acknowledges support from the Swedish Research Council, Reg. No. 637-2014-6863. M.A.H. was partially supported by NSF CAREER DMS-1255422. L.L.G. would like to thank the VolkswagenStiftung for the financial support through the Peter-Paul-Ewald Fellowship. E.I. thanks Leo Radzihovsky for fruitful discussions.

**Author contributions**

E.I. performed micromagnetic simulations. A.T. prepared the samples. A.H.R., T.-M.L., P.W.G., E.J., A.X.G., S.B., C.E.G., R.K., Z.C., D.J.H., T.C., L.L.G., K.H., H.O., W.F.S., G.L.D., G.C., M.C.H., S.C., M.S., A.K., A.V.K., T.R., J.S., and H.A.D. performed experiments. Z.F. and R.F.L.E. performed atomistic simulations. R.F.L.E. developed the numerical representation of Fe and Gd inhomogeneities in the atomistic model. Z.F., S.R., R.F.L.E., T.O. and R.W.C. analysed the atomistic data. T.-M.L. and D.H. analysed the experimental data. E.I., M.A.H., and T.J.S. fitted the data and analysed the micromagnetic simulations. All authors contributed to discussions, data analysis, and writing the manuscript.

**Competing financial interests**

The authors declare no competing financial interests.

|  | Sub-threshold | | Switching | |
| --- | --- | --- | --- | --- |
|  | *a* | *b* | *a* | *b* |
| Experiment | 0.81 ± 0.01 | 5.24 ± 0.14 | 0.76 ± 0.083 | 21.10 ± 4.39 |
| Atomistic simulations | 0.57 ± 0.005 | 9.89 ± 0.15 | 0.83 ± 0.02 | 10.88 ± 0.41 |
| Micromagnetic simulations | 1.08 ± 0.006 | 1.63 ± 0.03 | 1.04 ± 0.03 | 2.18 ± 0.27 |

**Table** 1**. Fitted parameters for the power law** $L(t) = bt^a$**.**

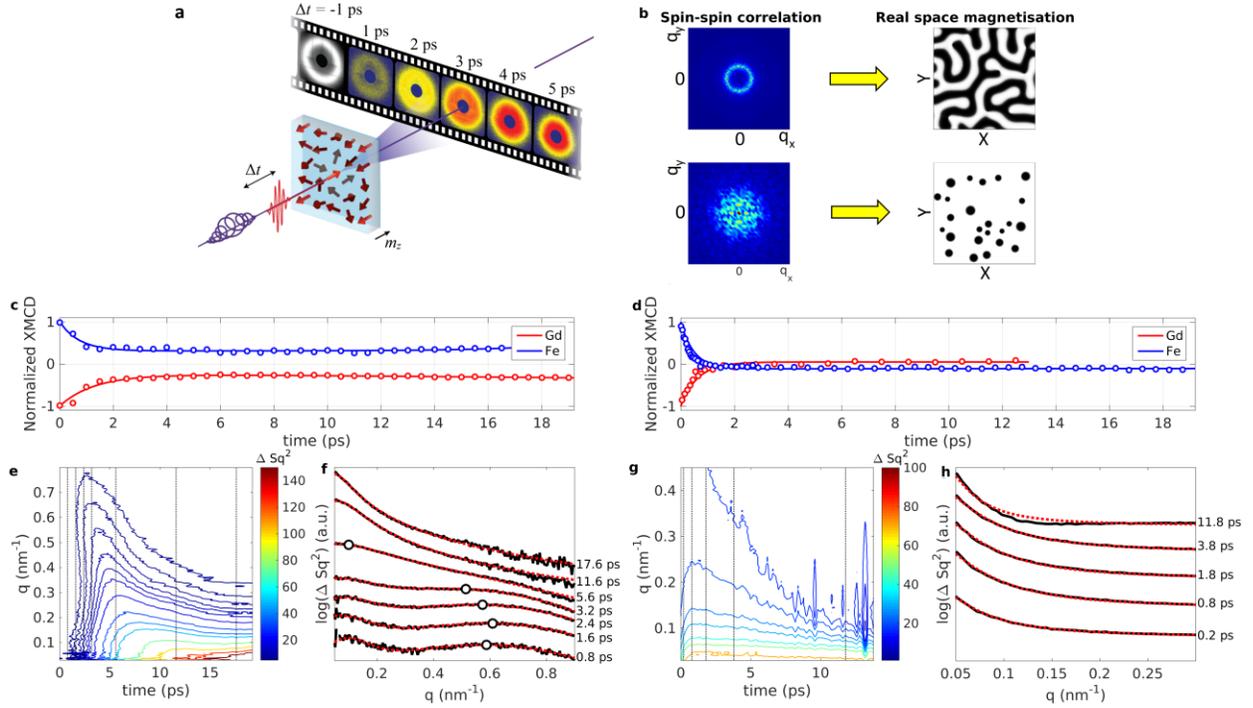

**Figure 1. Experimental setup and picosecond evolution of magnetisation dynamics. a** Schematic of the experimental setup. A femtosecond optical pulse randomises the spin degree of freedom and a subsequent circularly polarised X-ray pulse probes the perpendicular magnetisation, $m_z$, at a given delay, $\Delta t$. For each time delay, the two-dimensional X-ray scattering map is obtained, from which the spin-spin correlation function can be extracted. Numerical examples of the two-dimensional spin-spin correlation function and its associated spatial magnetisation are shown in **b**. XMCD data is shown in **c** for sub-threshold dynamics obtained in a 30 nm-thick sample subject to an absorbed fluence of 3.91 mJ/cm$^2$ and **d** for AOS obtained in a 20 nm-thick sample subject to an absorbed fluence of 4.39 mJ/cm$^2$. Solid lines are guides to the eye. **e** Contours of the azimuthally averaged spin-spin correlation function, $\Delta Sq^2$, for sub-threshold dynamics. For the time instances indicated by dotted vertical lines, lineouts are shown by black curves in **f** and are vertically shifted for clarity. Fits to the data with both Lorentzian and Gaussian components are shown by dashed red curves. The black circles indicate the peak position of the Gaussian component. **g** Contours of the azimuthally averaged spin-spin correlation function, $\Delta Sq^2$, for AOS. For the time instances indicated by dotted vertical lines, lineouts are shown by black curves in **h** and are also vertically shifted for clarity. Fits to the data with a Lorentzian lineshape are shown by dashed red curves.

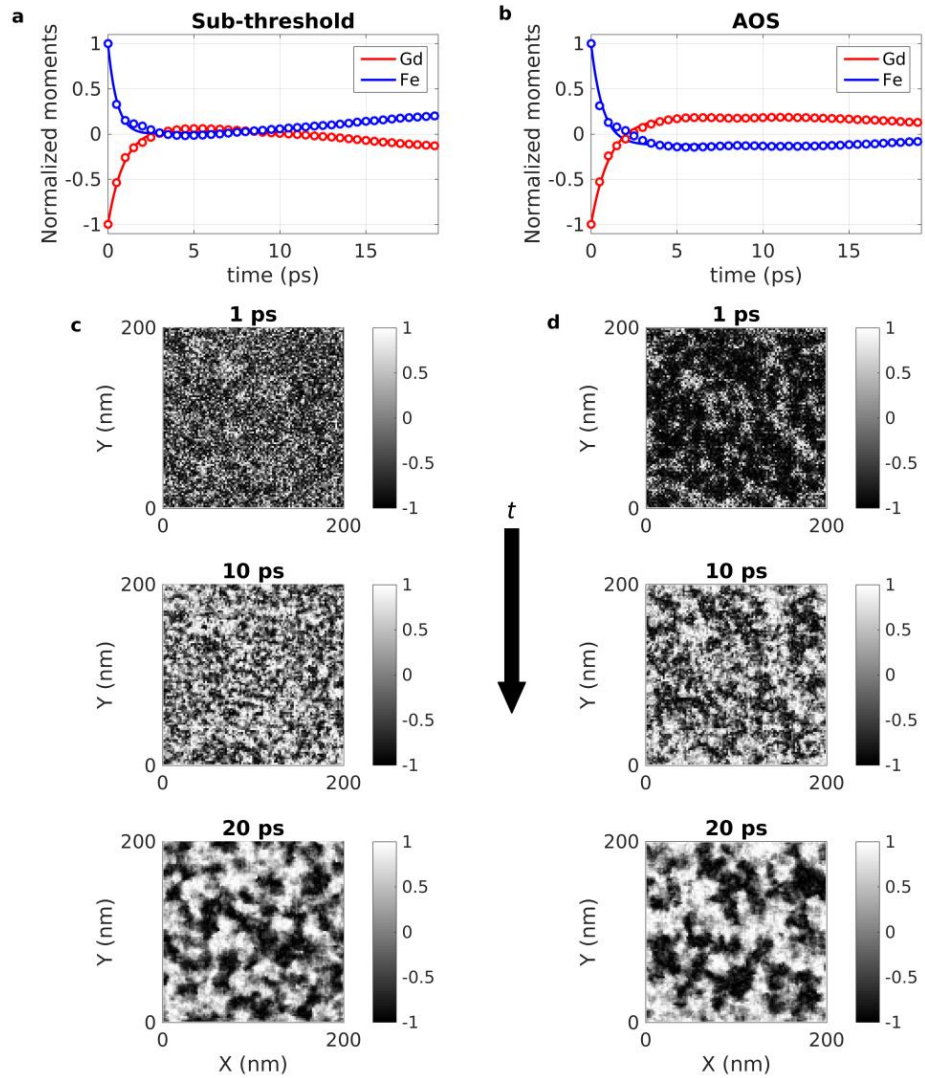

**Figure 2. Simulated magnetisation dynamics.** Normalized Gd and Fe average moments from atomistic simulations in the case of **a** sub-threshold dynamics obtained with a fluence of 10.7 mJ/cm$^2$, and **b** AOS obtained with a fluence of 11 mJ/cm$^2$. Snapshots of the perpendicular-to-plane magnetisation at 1 ps, 10 ps, and 20 ps for the case of **c** sub-threshold dynamics and **d** AOS. In both cases, the magnetisation exhibits coarsening of textures.

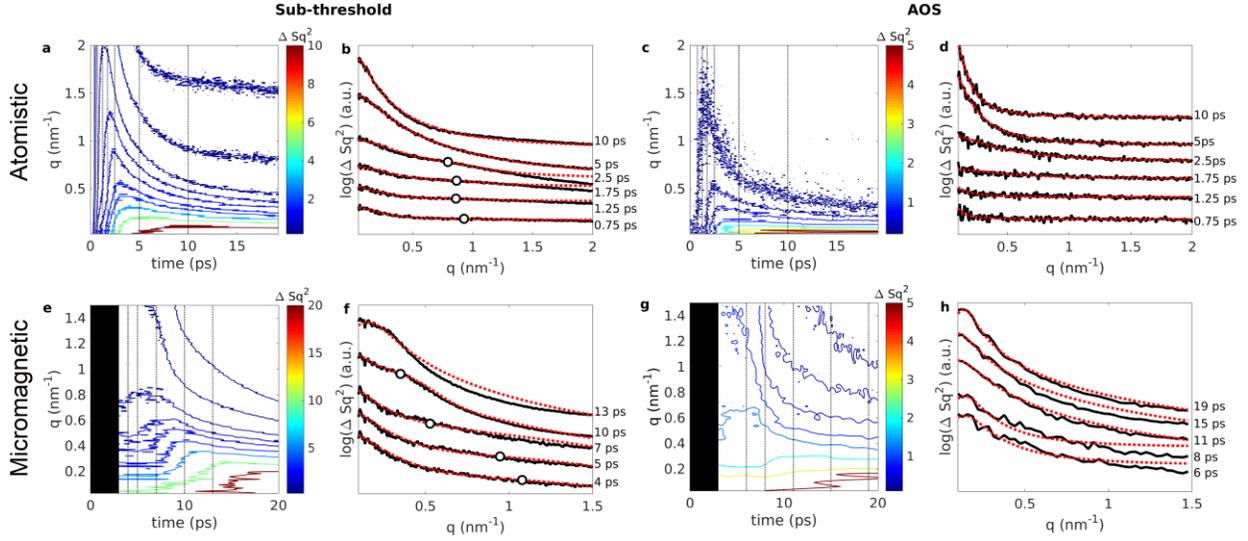

**Figure 3. Simulated spin-spin correlation functions. a** Contours of the azimuthally averaged spin-spin correlation function obtained from atomistic simulations. For the time instances indicated by dotted vertical lines, lineouts are shown by black curves in **b** and are vertically shifted for clarity. Fits using Lorentzian and Gaussian components are shown by read dashed lines. The peak position of the Gaussian component is shown by black circles. Equivalent plots for the case of AOS are shown in panels **c** and **d**. Fits to the lineouts in this case are obtained by using only a Lorentzian lineshape. For micromagnetic simulations seeded with an atomistic input at 3 ps, the azimuthally averaged spin-spin correlation function and corresponding lineouts and fits are shown in **e** and **f** for sub-threshold dynamics; and **g** and **f** for AOS.

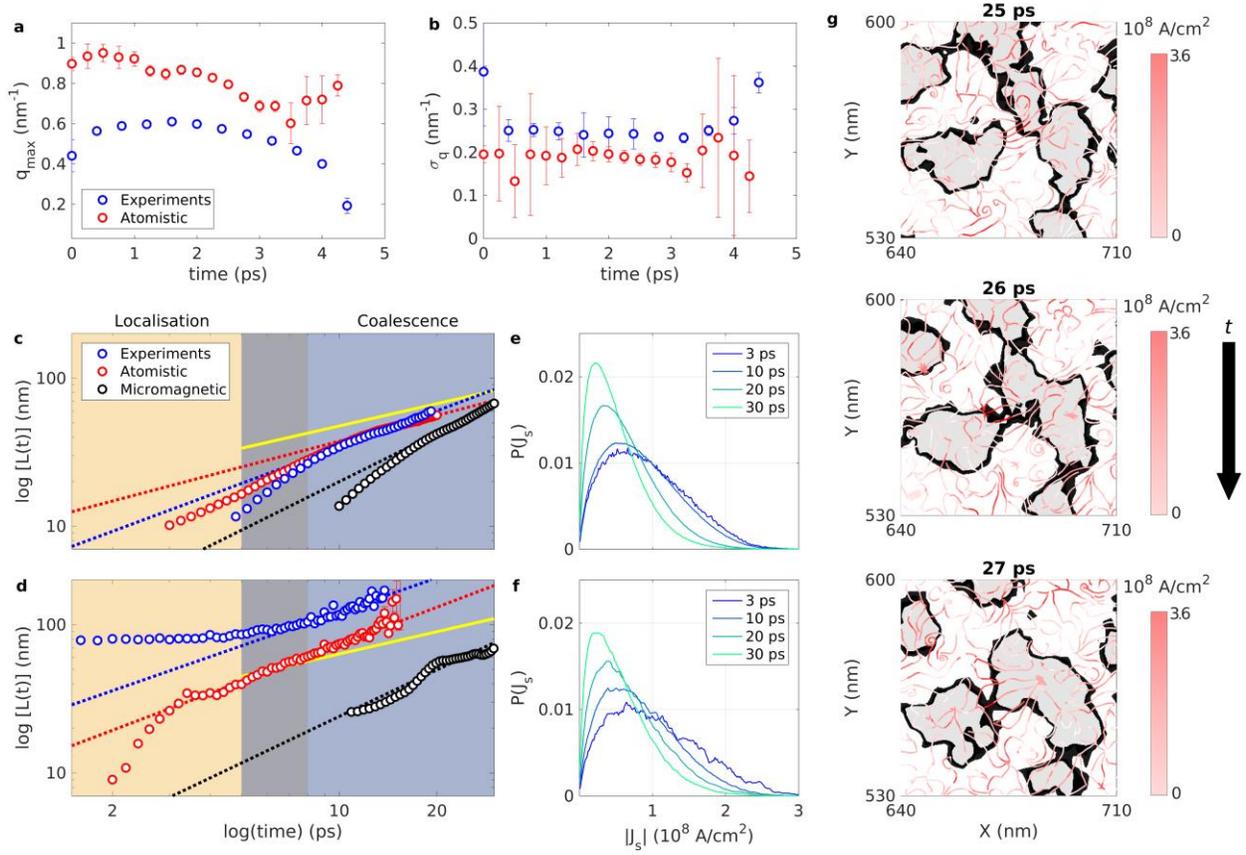

**Figure 4. Magnon localisation and coalescence. a** Peak position of the Gaussian feature from experiments (blue circles) and atomistic simulations (red circles) during magnon localisation. The corresponding peak width is shown in **b**. **c** Temporal evolution of the characteristic length scale in loglog scale for the case of sub-threshold dynamics obtained from Lorentzian fits of the experimental (blue circles), atomistic (red circles), and micromagnetic (black circles) data. Dotted lines with corresponding colour code are power-law fits. The yellow line indicates the Lifshitz-Cahn-Allen power law. The equivalent plot for the case of AOS is shown in **d**. The EFSC density probability distribution expressed as a 100% spin polarised charge current magnitude at selected time instances calculated from the micromagnetic simulations is shown in **e** for sub-threshold dynamics and **f** for AOS. **g** Example of magnon drop dynamics, including merging and break-up. The black areas represent magnon drop perimeters and the white and gray areas indicate that the perpendicular-to-plane magnetisation is parallel or antiparallel to the applied field. The red-shaded curves represent EFSCs flow with equivalent 100% spin polarised charge current magnitudes in the $10^8$ A/cm$^2$. The flow curves represent the streamlines in which perpendicular-to-plane angular momentum is transferred.

**Supplementary material: Spin-current-mediated rapid magnon localisation and coalescence after ultrafast optical pumping of ferrimagnetic alloys**

E. Iacocca[1,2,3], T-M. Liu[4], A. H. Reid[4], Z. Fu[5], S. Ruta[6], P. W. Granitzka[4], E. Jal[4], S. Bonetti[4], A. X. Gray[4], C. E. Graves[4], R. Kukreja[4], Z. Chen[4], D. J. Higley[4], T. Chase[4], L. Le Guyader[4,7], K. Hirsch[4], H. Ohldag[4], W. F. Schlotter[4], G. L. Dakovski[4], G. Coslovich[4], M. C. Hoffmann[4], S. Carron[4], A. Tsukamoto[8], M. Savoini[9], A. Kirilyuk[9], A. V. Kimel[9], Th. Rasing[9], J. Stöhr[4], R. F. L. Evans[6], T. Ostler[10,11], R. W. Chantrell[6], M. A. Hoefer[1], T. J. Silva[2], H. A. Dürr[4,12]

[1] Department of Applied Mathematics, University of Colorado, Boulder, CO 80309, USA
[2] National Institute of Standards and Technology, Boulder, CO 80305, USA
[3] Department of Physics, Division for Theoretical Physics, Chalmers University of Technology, Gothenburg 412 96, Sweden
[4] SLAC National Accelerator Laboratory, 2575 Sand Hill Road, Menlo Park, CA 94025, USA
[5] School of Physics, Science, and Engineering, Tongji University, Shanghai 200092, China
[6] Department of Physics, University of York, York YO10 5DD, UK.
[7] Spectroscopy & Coherent Scattering, European X-Ray Free-Electron Laser Facility GmbH, Holzkoppel 4, 22869 Schenefeld, Germany
[8] Department of Electronics and Computer Science, Nihon University, 7-24-1 Narashino-dai Funabashi, Chiba 274-8501, Japan
[9] Radboud University, Institute for Molecules and Materials, Heyendaalseweg 135, 6525 AJ Nijmegen, The Netherlands
[10] Université de Liège, Physique des Matériaux et Nanostructures, Liège B-4000 Sart Tilman, Belgium
[11] Faculty of Arts, Computing, Engineering and Sciences, Sheffield Hallam University, Howard Street, Sheffield, S1 1WB, UK
[12] Department of Physics and Astronomy, Uppsala University, Box 516, 751 20 Uppsala, Sweden

## S1. Short-time evolution of Gaussian feature

A Gaussian feature that corresponds to the magnetisation pattern seeded by the material chemical inhomogeneity is observed in the spin-spin correlation function for Gd. In Figure S1**a**, the data obtained for Gd in the sub-threshold case is shown as artificially shifted solid black curves from 0 ps (bottom lineout) to 4.8 ps (upper lineout). Fits with a Lorentzian and a Gaussian component are shown by dashed red curves. The peak position of the Gaussian component in time is shown by black circles. Whereas a Gaussian component can be fitted at 0 ps with some accuracy (when the sample is at thermal equilibrium), the feature is clearly seen only at the first measured delay after the femtosecond pulse. The corresponding evolution of the fitted Gaussian component is shown in Figure S1**b**. It is noteworthy that a Gaussian component appears even while the demagnetisation process is operative. Further theoretical work is required to disentangle the relative magnitudes between the randomisation of the spin degree of freedom due to coupling to the electronic and atomic thermal baths and the recovery of magnetic order mediated by the sample microstructure.

We note that fits obtained by utilising a Lorentzian line shape return similar metrics. An example of a Lorentzian line shape fit at t = 0.8 ps is shown in Figure S1**c** by a dashed blue curve. The fit is very similar to that obtained with a Gaussian line shape shown by a dashed red curve. However, we find as a general trend, that a Gaussian line shape returns smaller errors in the fitted quantities than a Lorentzian line shape.

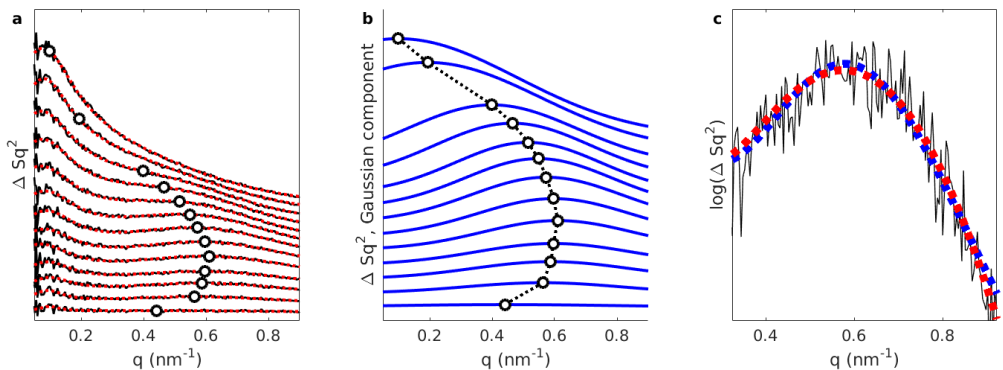

**Figure S**1**. Short time evolution of Gaussian component in the spin-spin correlation function. a** Lineouts of the spin-spin correlation function measured for Gd between 0 and 4.8 ps at a fluence of 3.91 mJ/cm$^2$ are shown by shifted solid black curves. Fits with Lorentzian and Gaussian components are shown by dashed red curves. **b** Gaussian component of the fitted lineouts. The peak position of the Gaussian component is shown by black circles in both panels. **c** Comparison of fits utilising a Lorentzian (dashed blue curve) and a Gaussian (dashed red curve) component.

## S2. Magnon coalescence for sub-threshold dynamics: Gd and Fe

Sub-threshold dynamics occur in our GdFeCo alloys for a range of fluences. X-ray scattering is measured simultaneously for Gd and Fe because of the technique's element specificity. For both elements and the absorbed laser fluences of 3.91 mJ/cm$^2$, 2.79 mJ/cm$^2$, and 1.39 mJ/cm$^2$, the contours of the azimuthally averaged spin-spin correlation function shown in the top row of Figure S2 exhibit similar qualitative features. The data for Fe has a lower signal-to-noise ratio but reliable fits to Lorentzian line shapes are achieved after 5 ps. The calculated characteristic length scale for each case is shown in the bottom row of Figure S2. Power-law fits can be obtained at long times for all cases, with parameters shown in each panel. The fact that modest fluences induce similar features in the spin-spin correlation function as well as evidence for growth suggests that the in-plane magnetisation may be highly randomised in all cases.

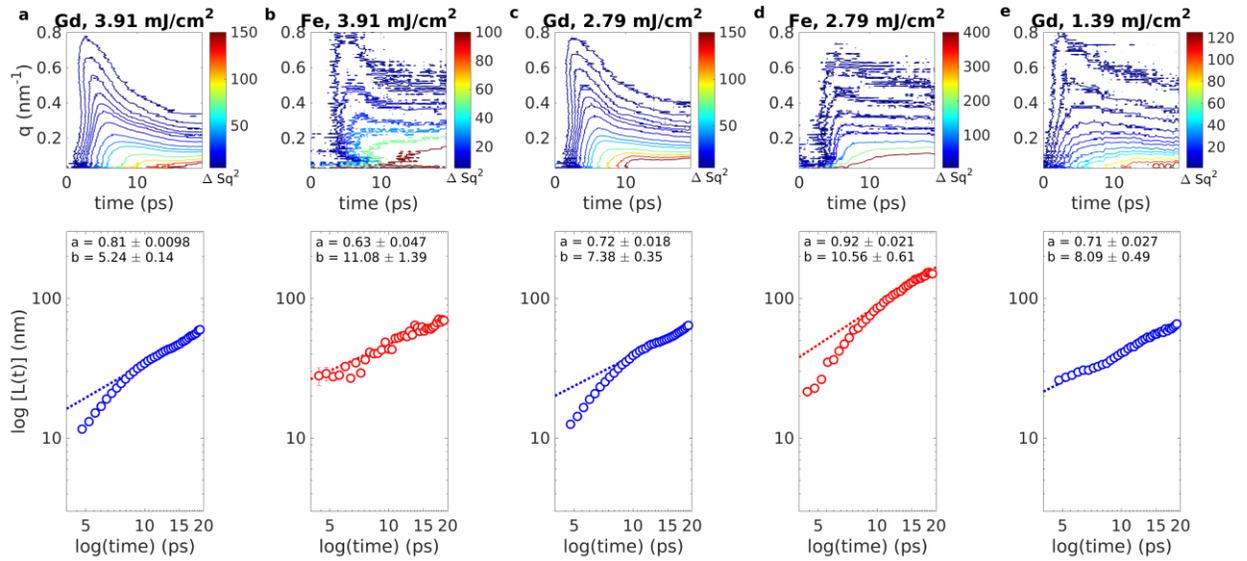

**Figure S2. Experimental data for Gd and Fe at several fluences.** Top row: contours of the azimuthally averaged spin-spin correlation function for the indicated element and fluence, namely, **a** Gd at 3.91 mJ/cm$^2$, **b** Fe at 3.91 mJ/cm$^2$, **c** Gd at 2.79 mJ/cm$^2$, **d** Fe at 2.79 mJ/cm$^2$, and **e** Gd at 1.39 mJ/cm$^2$. Bottom row: characteristic length scale calculated from Lorentzian fits to the azimuthally averaged spin-spin correlation function.

## S3. Micromagnetic exchange constant: average atomistic and micromagnetic dynamics

To obtain a multiscale model, the micromagnetic parameters for the GdFeCo alloy were chosen to match atomistic simulations. The saturation magnetisation, anisotropy constant, and damping can be directly obtained from atomistic simulations. The exchange constant is challenging to obtain because it requires an average on the element and spatially dependent Heisenberg exchange. The addition of inhomogeneity adds complexity to the spatial average calculation that leads to an imprecise determination of a micromagnetic exchange constant. To circumvent this problem, we utilised a numerical approach to estimate the micromagnetic exchange constant based on the qualitative behaviour of the perpendicular magnetisation, $<m_z>$. The goal was to choose an exchange constant such that the temporal evolution of $<m_z>$ calculated from micromagnetic simulations utilising atomistic magnetisation states as inputs at different times was both self-consistent, i.e., followed the same qualitative evolution, and consistent with atomistic simulations. The results obtained with an exchange constant $A = 1$ pJ/m are shown in Figure S3. Utilising atomistic spatial magnetisation as initial conditions at and after 3 ps, the micromagnetic simulations exhibit a slow evolution of $<m_z>$ that is qualitatively consistent between the different micromagnetic simulations, shown by circles, and agrees with the effective perpendicular magnetisation obtained from atomistic simulations, shown by a dashed black curve.

For the atomistic spatial magnetisation at 1 ps and 2 ps, a stark disagreement is observed. This occurs because of the predominantly switched average magnetisation at short times after the demagnetisation event. Note that while the dynamic behaviour is sub-threshold, the large magnetic moment of Gd relative to Fe leads to an average switched magnetisation in the multiscale modelling: micromagnetic simulations model a ferromagnet and, consequently, has no available physical mechanism to recover the short-range order based on the antiferromagnetic Gd-Fe exchange interaction. For the atomistic input magnetisation at 1 ps, the dominantly switched magnetisation translates into a large anisotropy energy that strives to relax the magnetisation towards the negative pole, i.e., $m_z = -1$. For the atomistic input magnetisation at 2 ps, the magnetisation is close to zero. While for ferrimagnets this implies average compensated moments, in micromagnetic simulations this implies that the saturation magnetisation is negligibly small and, consequently, the dynamics are extremely slow.

We emphasize that the choice of the exchange constant described here is not critical to model the qualitative features of magnon coalescence nor impacts the conclusions drawn in the main text.

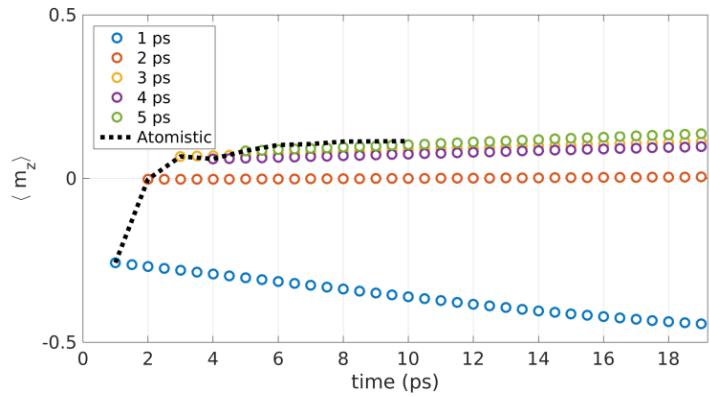

**Figure S3. Average magnetisation evolution from simulations.** Micromagnetic evolution of the average perpendicular magnetisation utilising atomistic spatial magnetisation as inputs at 1 ps, 2 ps, 3 ps, 4 ps, and 5 ps, shown by circles. The evolution of the effective perpendicular magnetisation from atomistic simulations is shown by a dashed black curve.

## S4. Multiscale simulations for sub-threshold dynamics as a function of $t_c$

The micromagnetic characteristic length scale growth presented in the main text was obtained by initialising the micromagnetic simulations with the atomistic spatial magnetisation at 3 ps. However, as shown in Figure S3, micromagnetic simulations exhibits a self-consistent behaviour utilising atomistic magnetisation states as inputs after 3 ps. The characteristic length scale growth calculated from Lorentzian fits to the azimuthally averaged spin-spin correlation function from micromagnetic simulations initialised with atomistic magnetisation states at times 3 ps, 4 ps, and 5 ps is shown in Figure S4. Despite a quantitative difference at short timescales (between 10 and 12 ps), the characteristic length scale growth converges, indicating that the multiscale simulations are accurately resolved.

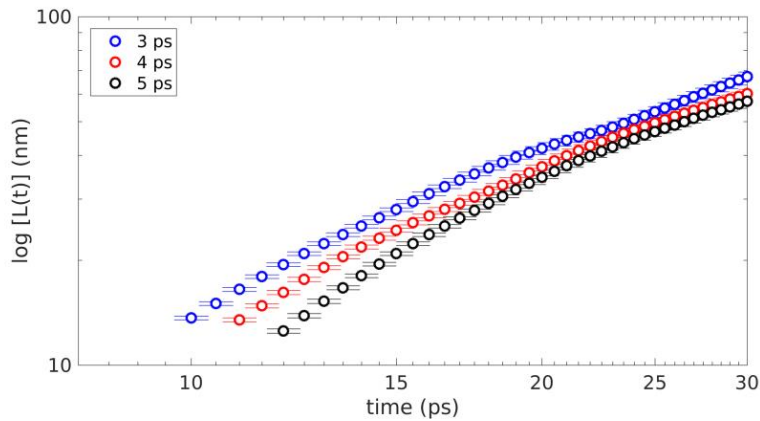

**Figure S4. Multiscale characteristic length scale growth.** Characteristic length scale growth calculated form Lorentzian fits to the spin-spin correlation function obtained from micromagnetic simulations initialised with atomistic magnetisation states at 3 ps (blue circles), 4 ps (red circles), and 5 ps (black circles).

## S5. Fast quench: micromagnetic simulations

Upon femtosecond heating, we have demonstrated that the magnetisation dynamics undergo localisation of magnons whereby the magnetisation rapidly forms a microscopic pattern mediated by the sample microstructure. This microscopic mechanism is resolved by atomistic simulations that accurately model the spin dynamics. However, it can be argued that the localisation of magnons is not a necessary process to develop localised textures. In fact, modulational instability provides a mechanism for magnons to localise in an ideal magnetic material with perpendicular magnetic anisotropy (PMA). Consequently, an ideal ferromagnet with PMA subject to quench should also exhibit nucleation of localised textures and subsequent magnon coalescence. To test this hypothesis, we perform micromagnetic simulations initialised with a random magnetisation distribution, i.e., a paramagnetic state, in an otherwise homogeneous magnetic system.

The resulting azimuthally averaged spin-spin correlation function is shown in Figure S5**a**. Remarkably, we observe the same qualitative features as in atomistic simulations and experiments. The main difference lies in the timescales. We emphasize that modulational instability expected to play a crucial role in magnon localisation for a homogeneous magnetic system is a mechanism that holds for small, long wavelength perturbations about a homogeneous state. A theory for modulational instability in the case of a randomised magnetisation is yet to be developed.

Lorentzian fits can be performed with good accuracy from 4 ps, as shown by the small errorbars in the Lorentzian peak position shown in Figure S5**b**. Notably, the peak position shifts but does not reaches zero during the simulated time of 40 ps in contrast to atomistic and multiscale simulations. The corresponding calculated characteristic length scale growth is shown in Figure S5**c**. A qualitatively similar growth is observed throughout the simulation and quantitatively agrees with the multiscale simulation beyond 20 ps. This indicates that the onset of magnon coalescence growth is independent of the mechanism that recovers magnetic order at short timescales, i.e., magnon localisation that is arguably unique to amorphous ferrimagnetic alloys. Instead, it is a feature of magnetic systems whose in-plane magnetisation component is strongly randomised.

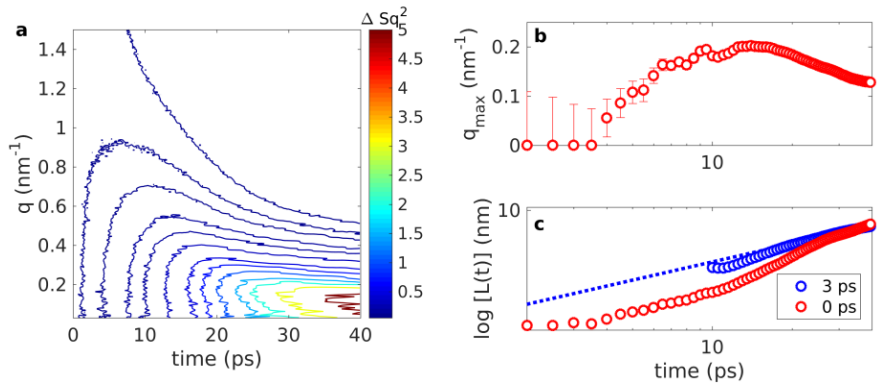

**Figure** S5**. Micromagnetic simulations starting with a randomised magnetisation. a** Azimuthally averaged spin-spin correlation. **b** Peak position and **c** calculated characteristic length scale growth from Lorentzian fits to the azimuthally averaged spin-spin correlation function. The power law fit in the coalescence regime is shown by a dashed blue line.